\documentclass[fleqn,usenatbib]{mnras_tex_edited}

\usepackage{amssymb,amsmath,verbatim,mathtools,needspace,enumitem,etoolbox,graphicx,physics,microtype,natbib,url,hyperref,mathrsfs,bm,ulem,lipsum}
\usepackage{orcidlink}
\usepackage{siunitx}

\renewcommand{\emph}[1]{\textit{#1}}

\DeclareRobustCommand{\VAN}[3]{#2}
\let\VANthebibliography\thebibliography
\def\thebibliography{\DeclareRobustCommand{\VAN}[3]{##3}\VANthebibliography}

\usepackage{graphicx,epsf, epsfig}
\usepackage{bm}
\usepackage{longtable,tabularx}
\usepackage[dvipsnames]{xcolor}
\usepackage{array}
\usepackage{multirow}
\usepackage{rotating,array}
\usepackage{dcolumn}
\usepackage{braket}
\usepackage{appendix}
\usepackage{comment}
\usepackage{siunitx}
\usepackage{adjustbox}

\definecolor{linkcolor}{rgb}{0.0,0.3,0.5}

\newcommand{\bham}{School of Physics and Astronomy \&	 Institute for Gravitational Wave Astronomy, University of Birmingham,\vspace{-0.05cm}\\$\;$Birmingham, B15 2TT, UK}

\def\apj{\rm{Astrophys.~J.}}
\def\apjl{\rm{Astrophys.~J.~ Lett.}}
\def\apjs{\rm{ApJS}}
\def\aap{\rm{Astron.~Astrophys.}}

\def\mnras{\rm{Mon.~Not.~R.~Astron.~Soc.}}
\def\prd{\rm{Phys.~Rev.~D}}
\def\prl{\rm{Phys.~Rev.~Lett.}}

\def\pasp{\rm{PASP}}

\def\nat{\rm{Nature}}

\def\physrep{\rm{Phys.~Rep.}}

\title[High spin and kilonova emission]{Mechanisms for high spin in black-hole neutron-star binaries and kilonova emission: inheritance and accretion}

\author[N. Steinle, B. P. Gompertz, and M. Nicholl]{Nathan Steinle$\,$\orcidlink{0000-0003-0658-402X}$^{1}$\medskip\thanks{Contact e-mail: \href{mailto:nsteinle@star.sr.bham.ac.uk}{nsteinle@star.sr.bham.ac.uk}},
Benjamin P. Gompertz$\,$\orcidlink{0000-0002-5826-0548}$^{1}$\medskip,
Matt Nicholl$\,$\orcidlink{0000-0002-2555-3192}$^{1}$\medskip
\\
$^{1}$\bham
}

\date{Accepted XXX. Received YYY; in original form ZZZ}

\pubyear{\the\year{}}

\begin{document}
\label{firstpage}
\pagerange{\pageref{firstpage}--\pageref{lastpage}}
\maketitle

\begin{abstract}
A black-hole neutron-star binary merger can lead to an electromagnetic counterpart called a kilonova if the neutron star is disrupted prior to merger. The observability of a kilonova depends on the amount of neutron star ejecta, which is sensitive to the aligned component of the black hole spin. We explore the dependence of the ejected mass on two main mechanisms that provide high black hole spin in isolated stellar binaries. When the black hole inherits a high spin from a Wolf-Rayet star that was born with least $\sim 10\%$ of its breakup spin under weak stellar core-envelope coupling, relevant for all formation pathways, the median of the ejected mass is $\gtrsim 10^{-2}$ M$_{\odot}$. Though only possible for certain formation pathways, similar ejected mass results when the black hole accretes $\gtrsim 20\%$ of its companion's envelope to gain a high spin. Together, these signatures suggest that a population analysis of black-hole neutron-star binary mergers with observed kilonovae may help distinguish between mechanisms for spin and possible formation pathways. We show that these kilonovae will be difficult to detect with current capabilities, but that future facilities, such as the Vera Rubin Observatory, can do so even if the aligned dimensionless spin of the black hole is as low as $\sim 0.2$. Our model predicts kilonovae as bright as $M_i \sim -14.5$ for an aligned black hole spin of $\sim 0.9$ and mass ratio $Q = 3.6$.
\end{abstract}




\section{Introduction}
\label{sec:Intro}
Although the majority of observed gravitational waves are sourced by black-hole binary mergers, the LIGO and Virgo collaborations reported the detection of seven events that are consistent with a new class of compact binary: black-hole neutron-star (BHNS) binary mergers.
For examples, assuming uninformative priors, the masses of GW200115 and GW200105 are $8.9^{+1.2}_{-1.5}$ M$_{\odot}$ and $1.9^{+0.3}_{-0.2}$ M$_{\odot}$, and $5.7^{+1.8}_{-2.1}$ M$_{\odot}$ and $1.5^{+0.7}_{-0.3}$ M$_{\odot}$, respectively, at the 90\% credible level \citep{LIGO2021BHNS}. The spin of the black hole in GW200115 is not tightly constrained but may be misaligned as it is inferred to have a component below the orbital plane at 88\% probability, while the dimensionless spin magnitude of the black hole in GW200105 is likely $< 0.23$ and its direction is unconstrained. Observations with future ground-based detectors may uncover more BHNS binaries and shed light onto their peculiar properties \citep{Brown2021}.

If a neutron star (NS) is tidally disrupted by its black hole (BH) companion rather than directly plunging beyond its event horizon \citep{Foucart2020}, $\gamma$-ray emission in the form of a short gamma-ray burst (GRB) may result from accretion onto the remnant stellar-mass BH \citep{Rosswog05,Lee2007,Paschalidis15}, and radioactive decay in neutron-rich ejecta may produce a roughly isotropic optical/infra-red emission known as a kilonova \citep{Li98,Roberts2011,Metzger2012,Barnes13,Metzger17}. The observability of an electromagnetic counterpart depends on the amount of mass ejected prior to merger. This is sensitive to the binary mass ratio \citep{Kyutoku2011,Etienne2009,Foucart2012b}, the compactness of the NS (i.e., its mass and radius) \citep{Kyutoku2010,Duez2010,Kyutoku2011}, and the aligned spin component of the BH  \citep{Etienne2009,Foucart2011,Foucart2012b,Foucart2013,Kawaguchi15} because the radius of the innermost stable circular orbit is smaller for higher prograde aligned spin. Although optical follow-up was not completely comprehensive, e.g., only $\approx 50\%$ of the sky location probabilities were searched by the \emph{Zwicky} Transient Facility \citep{Anand2021}, no electromagnetic counterparts were observed for either BHNS event detected by LIGO/Virgo consistent with theoretical expectations from their measured spins and mass ratios \citep{Gompertz2022}.

Theoretically, BHNSs can form in two broad scenarios: the dynamical channel where the compact binary forms in a dense stellar cluster \citep{Benacquista2013}, and the isolated channel where an isolated stellar binary forms into the compact binary through the various stages of binary evolution \citep{Postnov2014}. Although both channels of formation may explain the origin of the LIGO/Virgo population of presently known binary BHs \citep[for a review, see e.g.][]{Mapelli2020}, the dynamical channel is expected to produce a substantially smaller number of merging BHNS binaries \citep{Clausen2013,Ye2020} compared to what is estimated from current LIGO/Virgo observations, and possibly no counterparts as the mass of the BH tends to be larger \citep{Sedda2020}. The merger rates of isolated BHNS binaries are highly uncertain due to the uncertainties of stellar binary evolution, galactic star-formation history, and cosmic evolution of the metallicity dependence of star-forming regions \citep[e.g.][]{Dominik2015,Giacobbo2018,Belczynski2020,Broekgaarden2021}. 

Population synthesis studies of merger rates of isolated BHNSs typically find that the vast majority of binaries will not result in observable electromagnetic counterparts \citep[e.g.][]{Zhu2021,Fragione2021}, although the spins of BHs remain uncertain theoretically, see e.g., \citep{Belczynski2020}, and observationally \citep{Miller2015}. The fraction of BHNSs that yield significant ejecta can be sensitive to the assumptions that are employed, especially assumptions regarding the BH spin. \citet{Drozda2020} found this fraction to be $\lesssim 20\%$ when core-envelope coupling of their stellar progenitors is sufficiently weak to provide a high dimensionless BH spin magnitude, i.e., $\chi = 0.9$. A similar fraction of binaries is reported under the assumption that the BH spin magnitude is $\chi = 0.5$ \citep{Broekgaarden2021}. 
The efficiency (strength) of angular momentum transport (core-envelope spin coupling) is an uncertain aspect of high-mass stellar evolution theory \citep{Maeder2012,Meynet2013,Belczynski2020} and is not well-constrained observationally \citep{Bowman2020}. If it is efficient implying the core of the stellar progenitor is born with negligible spin, \citet{Belczynski2020} pointed out that detection of an electromagnetic counterpart of a BHNS binary merger could imply the BH experienced significant accretion, its progenitor was tidally synchronized, or it experienced repeated mergers in dynamical formation. On the other hand, if it is inefficient detection of a counterpart may also imply the BH inherited high spin from its progenitor. 
Accretion in stable mass transfer is usually considered in population synthesis studies but has not been investigated as a mechanism of high BH spin for significant ejected mass. Similarly, a systematic study of the dependence of ejected mass on inherited spin from weak core-envelope coupling has not been explored. 

Motivated by these previous studies, we explore the dependence of the ejected mass of BHNS binaries in two evolutionary pathways of the model of isolated formation of \citet{Steinle2021}. This model simplified binary stellar evolution in order to parametrize various processes that are pertinent for evolving the binary spin magnitudes and directions. We focus on two mechanisms by which the BH obtains a high spin magnitude. The first is inheritance of natal spin via weak stellar core-envelope coupling of the stellar progenitor. Instead of assuming a single value of the inherited BH spin, e.g., as was done in \citet{Drozda2020}, we take this a step further by parametrizing the natal spin of the progenitor Wolf-Rayet star with the fraction, $f_{\rm B}$, of its maximum breakup spin. The second mechanism is accretion of a fraction, $f_{\rm a}$, of the donor's envelope that is accreted onto the BH during stable mass transfer. 
We use the formula of \citet{Foucart2018} to determine whether the NS is tidally disrupted allowing us to parametrize the ejected mass of the NS with the fractions $f_{\rm B}$ and $f_{\rm a}$. We do not attempt to compute the merger rate of our BHNS distributions since it would require the use of population synthesis. As electromagnetic observations of a kilonova from a BHNS would provide estimates of the ejected mass of the NS, it is crucial to understand how the uncertainties of binary stellar evolution can lead to BHNSs with large BH spins. While we cannot forecast reliable population statistics for BHNSs and the likelihood for kilonova, instead we attempt to identify key signatures of the uncertainties of BH spin-up processes and their impact on the possible kilonova emission. We expect that these signatures will be useful for statistical analyses of BHNSs with observed gravitational signals and kilonova counterparts.

This paper is organized as follows: in section \ref{sec:Meth} we detail our model of BHNS formation, NS tidal disruption, and counterpart light curves; in section \ref{sec:Results} we demonstrate the dependence of the ejected mass and the corresponding light curves on the mechanisms for obtaining high BH spin magnitude; and we conclude with a summary and discussion of implications in section \ref{sec:Disc}.

\section{Methodology}
\label{sec:Meth}

\subsection{Black-hole neutron-star binary formation}
\label{subsec:Formation}
A zero-age main sequence (ZAMS) binary star is initialized at the binary separation $a_{\rm ZAMS}$ with metallicity $Z$, and with masses $m_{1, \rm ZAMS}$ of the primary star and $m_{2, \rm ZAMS}$ of the secondary star such that the ZAMS mass ratio is $q_{\rm ZAMS} = m_{2, \rm ZAMS}/m_{1, \rm ZAMS} \leq 1$. For a detailed description of this model, see \citet{Steinle2021}, and for a detailed review of the physics of the isolated channel, see e.g., \citet{Postnov2014}.

\begin{figure}
\centering
\includegraphics[width=0.48\textwidth]{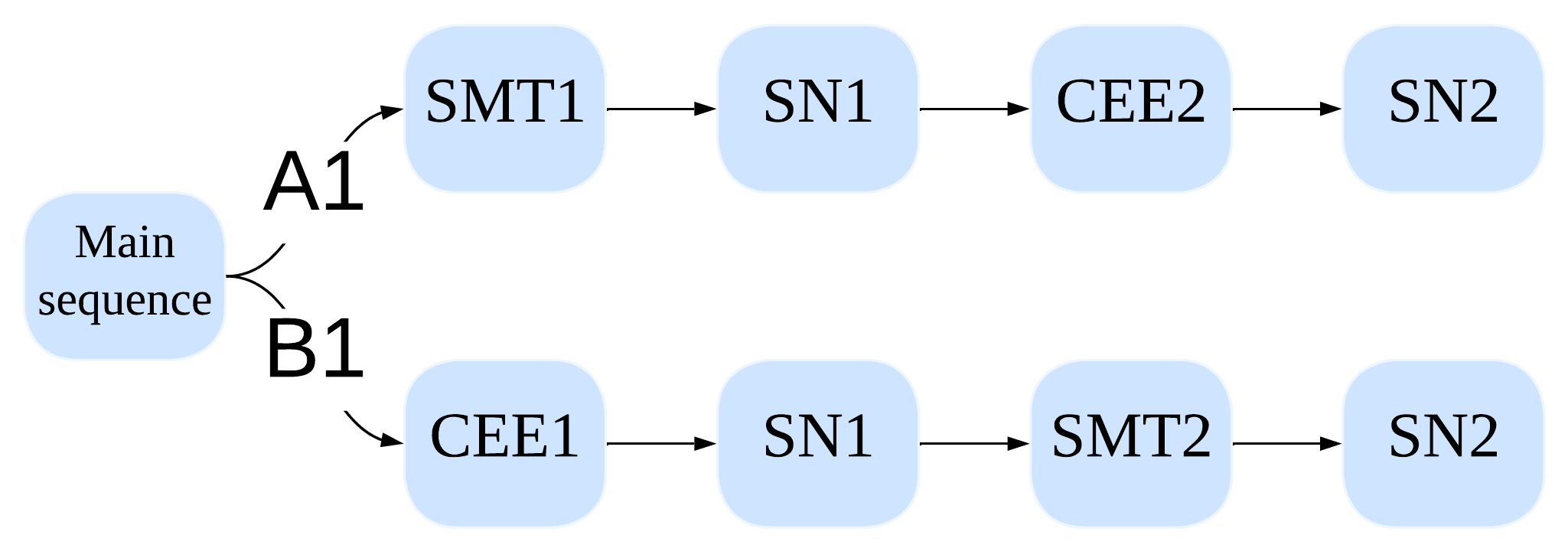}
\caption{A diagram of the two evolutionary pathways of black-hole neutron-star binary formation for which we present results in Section~\ref{sec:Results}. Stellar binaries are initialized on the main sequence and evolve in Pathway A1 when the initially more massive star undergoes stable mass transfer (SMT1) and the initially less massive star undergoes common envelope evolution (CEE2). A stellar binary evolves in Pathway B1 when the opposite order of events occurs, i.e., the initially more massive and less massive stars undergo CEE1 and SMT2, respectively.} \label{F:Diagram}
\end{figure}

Numerous astrophysical processes of isolated binary evolution are parametrized. Roche lobe overflow (RLOF) initiates mass transfer either as a phase of common-envelope evolution (CEE), which drastically shrinks the binary separation, or stable mass transfer (SMT), where the companion gains mass and spin angular momentum. The donor completely loses its envelope in mass transfer and its core emerges as a Wolf-Rayet (WR) star.
Rather than determining the stability of mass transfer via a critical mass ratio, as is typically done in population synthesis models, this is simplified into two evolutionary idealizations: Scenario A occurs when the primary undergoes SMT and the secondary undergoes CEE, while conversely so in Scenario B. Although this definition is independent of the ZAMS masses, the stellar lifetimes are mass dependent implying that both Scenarios admit two possible evolutionary pathways. Pathways A1 and B1 (A2 and B2) occur when the secondary star undergoes RLOF after (before) the core collapse supernova (SN) of the primary star.
We only present results for A1 and B1, which are depicted in Figure~\ref{F:Diagram}, because the boundary between pathways A1 and A2 (or equivalently, B1 and B2) defined by the mass ratio is large, i.e., $q_{\rm ZAMS} \approx 1$, for the total masses considered here, e.g., see Fig. 14 of \citet{Steinle2021}. 
Additionally, equal mass binaries are unlikely to form BHNSs as either binary BH or binary NS formation is more likely.

To form BHNS binaries, rather than BH binaries, we modify the model of \cite{Steinle2021}. Most importantly, we examine stars with lower ZAMS mass, i.e., 13~M$_{\odot} \leq m_{\rm ZAMS} \leq 25$~M$_{\odot}$. These stars may result in either NSs or BHs depending on the amount of fallback accretion onto the proto-NS during core-collapse formation.
We use the \texttt{StarTrack} implementation of the rapid energy-expenditure mechanism for the SN explosion, i.e., Eq.'s (10-14) of \citet{Fryer2012}. This provides the mass of the compact object and the fallback parameter $f_{\rm fb}$ which determines the fraction of material that falls onto the collapsing core after it was ejected during the SN. The fraction $f_{\rm fb}$ monotonically increases with increasing initial mass leading to larger compact remnant masses. This more physically motivated prescription produces a smooth transition across the uncertain parameter space between the \citet{Hurley2000} NS model (i.e., their Eq. (92)) and the BH model of \citet{Steinle2021}. We assume a mass boundary of $m = 2.5$ M$_{\odot}$ between NS and BH formation, as in \cite{Fryer2012}. Fallback accretion suppresses the natal kick imparted on the compact object that forms according to $v_{\rm k,fb} = (1 - f_{\rm fb})v_{\rm k}$ \citep{Fryer2012}, where BHs that form from stars with ZAMS masses $\gtrsim 23$ M$_{\odot}$ are assumed to experience complete fallback, i.e., $f_{\rm fb} = 1$, and do not experience a natal kick \citep{Heger2003}. The natal kick velocity magnitude $v_{\rm k}$ is drawn from a Maxwellian distribution with dispersion $\sigma$, and the natal kick velocity direction is spatially isotropic. A smaller value of $\sigma$ is required in Pathway A1 than in B1 to avoid unbinding too many binaries, as the primary SN (SN1) occurs before CEE has decreased the orbital separation. Stellar winds decrease the mass and spin of a star throughout its life. As we are interested in mechanisms that provide high BH spin, we only consider low ZAMS metallicity, i.e. $Z = 2\times10^{-4}$, where winds negligibly affect the mass and spin of the stellar progenitor (see e.g., \citet{Vink2001}).

Prior to core collapse, the WR star can experience tides from its companion. The tidal torque is a strong function of the binary separation $a$, i.e., the synchronization timescale $t_{\rm sync} \sim \left(a/{\rm R_{\odot}}\right)^{17/2}\left(m/{\rm M_{\odot}}\right)^{-7.54}$ where $m$ is the WR mass. This implies that tides are effective at producing high spin magnitudes after CEE. Tidal synchronization and alignment would seem to be a natural mechanism for producing significant ejected mass as the ejected mass is a strong function of the aligned component of the BH spin magnitude. However, tides in Pathway A1 are only effective on the secondary WR star as the primary BH forms before CEE occurs, and thus will not produce observable counterparts unless SMT onto the secondary main sequence star is sufficient to cause a mass-ratio reversal. \cite{Broekgaarden2021} find that highly conservative mass transfer may reverse the binary mass ratio to allow a tidally spun-up secondary star to form a highly spinning BH. Consistent with their results, we find this is realizable in Pathway A1 only for highly conservative mass transfer in a narrow region of parameter space, i.e. $q_{\rm ZAMS} > 0.9$, and therefore we do not explore this in detail here. In Pathway A2, tides can affect both WR stars potentially allowing for a high BH spin, but this requires fine-tuning to ensure that the secondary is still not too massive to form a BH, e.g., $q_{\rm ZAMS} \sim 0.95$. If tidal interactions were to produce a highly spinning and aligned WR star such a system may yield significant ejected mass \citep{Hu2022}.

Given the difficulties with tidal spin-up, we focus on two alternative mechanisms that may provide high BH spin: i) the BH inherits a high spin from weak core-envelope coupling of the stellar progenitors (relevant in pathways A1 and B1), and ii) the BH gains a high spin magnitude from accretion during SMT (relevant in Pathway B1).

A high dimensionless BH spin magnitude can be inherited in both pathways via minimal core-envelope coupling if its WR progenitor has a sufficiently large initial spin, $\chi_{\rm 0} = f_{\rm B} \chi_{\rm B}$, which is parametrized by the fraction $f_{\rm B}$ of the dimensionless breakup spin, defined as
\begin{align} \label{E:Break}
\chi_{\rm B}  = \frac{c |\bf{S}_{\rm B}|}{G m^2} = \frac{c r_{\rm g}^2 R^2 \Omega_{B}}{G m} = r_{\rm g}^2 \left( \frac{c^2R}{Gm} \right)^{1/2},
\end{align}
where $c$ is the speed of light, $G$ is the gravitational constant, $m$ is the mass of the WR star, $R$ is the WR stellar radius (see Eq.~(78) of \citet{Hurley2000}), $r_{\rm g}$ is the WR radius of gyration, and $\Omega_{\rm B}$ is the breakup angular frequency. For WR stars with $r_{\rm g}^2 = 0.075$, $\chi_{\rm B} \sim 15$ for $m \sim 10$ M$_{\odot}$. In the opposite extreme of maximal core-envelope coupling, angular momentum is efficiently transferred from the stellar progenitor's core to its envelope which is lost in mass transfer. This spin-down is modeled isotropically (see Eq.~(6) of \citet{Steinle2021}) and produces a natal WR dimensionless spin $\chi_0 \sim 0.001$ for $Z = 2\times10^{-4}$.

Accretion during SMT can result in a highly spinning primary BH in Pathway B1 depending on the fraction $f_{\rm a}$ of gas that is accreted. The increase in its mass $m_{\rm BH}$ and dimensionless spin $\chi$ per unit of accreted rest mass are given by,
\begin{subequations}
\begin{align}
\frac{dm_{\rm BH}}{dm} &= E(\chi)~,  \label{E:BHAcc1} \\ 
\frac{d\chi}{dm} &= \frac{L(\chi)}{m_{\rm BH}^2} - \frac{2\chi E(\chi)}{m_{\rm BH}}~, \label{E:BHAcc2}
\end{align}
\end{subequations}
where $E(\chi)$ and $L(\chi)$ are the specific energy and orbital angular momentum of massive particles at the (prograde) innermost stable circular orbit (ISCO) of the Kerr BH \cite{Bardeen1972}. We allow for super-Eddington accretion as in \citet{Steinle2021} (see their Appendix B.2). Although the secondary star accretes on the main sequence in Pathway A1, this accretion is ineffective at yielding a highly spinning BH as any spin that is gained is not inherited by the core under minimal core-envelope coupling or is dissipated under maximal core-envelope coupling during mass transfer.

\subsection{The ejected mass of a tidally disrupted neutron star}
\label{subsec:Ejecta}
Near the end of the BHNS binary inspiral, the NS can be tidally disrupted by its BH companion. A simple criterion for whether this produces an observable electromagnetic signal can be estimated by comparing the separation, $r_{\rm tid}$, at which tidal disruption occurs with the radius, $R_{\rm ISCO}$, of the innermost stable circular orbit (ISCO) of the BH. Ignoring general relativistic effects, the tidal disruption separation $r_{\rm tid}$ can be approximated by balancing the gravitational acceleration due to the NS, $\sim m_{\rm NS}/R_{\rm NS}^2$, with the tidal acceleration due to the BH, $\sim (m_{\rm BH}/r_{\rm tid}^3)R_{\rm NS}$, as $r_{\rm tid} \sim R_{\rm NS}(m_{\rm BH}/m_{\rm NS})^{1/3}$ \citep{Rees1988}. For a Kerr BH with dimensionless spin $\chi \equiv cS/Gm_{\rm BH}^2$ where $S$ is the magnitude of the spin angular momentum, $R_{\rm ISCO}$ is given by \citet{Bardeen1972} and depends sensitively on $\chi$. Tidal disruption is preceeded by mass-shedding of the outer layers of the NS at separations that are large compared to the ISCO of the BH when the tidal force exerted by the BH overcomes the self-gravity of the NS. As such, the separation at which mass-shedding begins is much larger than $r_{\rm tid}$, which is larger than $R_{\rm ISCO}$ when the NS is tidally disrupted. However, mass-shedding does not guarantee tidal disruption, and if the NS plunges into the BH before being tidally disrupted then only a low-mass accretion disk may form and an observable electromagnetic counterpart is very unlikely, see e.g., \citep{Kyutoku2021}.

The tidal disruption criterion, and the computation of the amount of ejected mass, is more accurately determined by fits to results of numerical relativity simulations. These typically quantify the criterion in terms of the BHNS binary mass ratio $Q \equiv m_{\rm BH}/m_{\rm NS} \geq 1$, $\hat{R}_{\rm ISCO} = R_{\rm ISCO}/m_{\rm BH}$ which depends on the aligned component of the BH spin, and the NS compactness $C_{\rm NS} = Gm_{\rm NS}/(R_{\rm NS}c^2)$. We use the formula of \citet{Foucart2018} to determine whether the NS is tidally disrupted and to compute the corresponding ejected mass,
\begin{align}\label{E:EjectedMass}
\begin{aligned}
    m_{\rm ejecta} = \left[ {\rm Max}\left( \alpha\frac{1 - 2C_{\rm NS}}{\eta^{1/3}} - \beta\hat{R}_{\rm ISCO}\frac{C_{\rm NS}}{\eta} + \gamma , 0 \right) \right]^\delta  m_{\rm NS}\,
\end{aligned}
\end{align}
where $\eta = Q/(1 + Q)^2$ is the symmetric mass ratio which enforces invariance of a change of labels of the NS and BH, $\alpha = 0.406,~\beta = 0.139,~\gamma = 0.255$, and $\delta = 1.761$ are constants derived from fitting the above model to 75 numerical relativity simulations \citep{Foucart2018}, and $m_{\rm NS}$ is in units of M$_{\odot}$. Eq~(\ref{E:EjectedMass}) is zero if the NS is not tidally disrupted, and is nonzero if the NS is tidally disrupted. The ejected mass is largest for highly spinning and aligned BH spin magnitudes as $R_{\rm ISCO}$ is smallest for a (prograde) maximally spinning BH with small mass. A larger NS mass or a smaller NS radius will result in a larger compactness $C_{\rm NS}$ and a smaller ejected mass.

We parametrize the ejected mass of BHNSs by the initial spin of the Wolf-Rayet progenitor star as a fraction of its breakup spin $f_{\rm B}$, and by the fraction of the donor's envelope $f_{\rm a}$ that is accreted in stable mass transfer. We also explore the dependence of $m_{\rm ejecta}$ on the binary component masses, $m_{1,2, \rm ZAMS}$, and the strength of the natal kicks at formation $\sigma$. Significant spin-orbit misalignments suppress $m_{\rm ejecta}$ by diminishing the aligned component of the BH spin. The effect of eccentricity is not considered in Eq.~(\ref{E:EjectedMass}), but the supernova of the secondary star (SN2) can introduce eccentricity into the binary system. We compute the time to coalescence \citep{Peters1964} of our BHNS binaries with their semi-major axes and eccentricities after SN2, and only compute $m_{\rm ejecta}$ for circularized binaries that merge within the age of the Universe. 
Although Eq.~(\ref{E:EjectedMass}) is calibrated to a set of numerical simulations with dimensionless BH spin $\leq 0.97$, it is possible for the BHs in our BHNS binaries to obtain spins $> 0.97$. Despite this we use Eq.~(\ref{E:EjectedMass}) which introduces a systematic error in $m_{\rm ejecta}$ for BHNSs with nearly maximal BH spin.

\subsection{Electromagnetic counterparts}
Having calculated the ejected mass for our BHNS binaries, we can take our analysis a step further by predicting their electromagnetic counterparts. Our BHNS kilonova model is from Gompertz et al. (in prep), where the full details will be presented. Here, we summarise the physics required to convert ejected mass into kilonova light curves. We note that although this is a somewhat simplified analytic model rather than a full kilonova simulation, the advantage of this approach is that we can quickly generate a reasonable estimate of the kilonova light curve for arbitrary binary parameters. This allows us to directly connect this model to the end products of our binary evolution calculations. We divide the total ejected mass $m_{\rm ejecta}$ in Eq~(\ref{E:EjectedMass}) into two post-merger components: unbound dynamical ejecta $m_{\rm dyn}$ \citep{Kruger20} and bound disc mass $m_{\rm disc} = m_{\rm ejecta} - m_{\rm dyn}$. The average velocity of the dynamical ejecta is determined from the fitting function of \citet{Kawaguchi16}, who found it has a linear relation with Q in numerical relativity simulations. We model the dynamical ejecta with a grey opacity of $\kappa_{\rm dyn} = 10$\,cm$^2$\,g$^{-1}$ \citep{Tanaka13,Kasen17,Tanaka20}.

Simulations show that winds will be driven from the surface of the disc by viscous heating and nuclear recombination \citep[e.g.][]{Fernandez13,Fernandez15,Just15,Fernandez20,Fujibayashi20,Kyutoku2021}. We parameterise the mass of this thermal wind as a fraction of the disc mass $m_{\rm therm} = \xi m_{\rm disc}$, where $\xi$ is a function of Q \citep{Raaijmakers21}, and assume a velocity $v_{\rm therm} = 0.034c$ \citep[cf.][]{Fernandez20}. The average electron fraction ($Y_e$) of the thermally-driven outflow is expected to be in the range $0.25 \leq Y_e \leq 0.35$ \citep[e.g.][]{Foucart15,Fernandez20,Fujibayashi20}, with $> 50$ per cent of the outflow expected to possess a Lanthanide + Actinide fraction $X_{(La+Ac)} < 10^{-4}$ \citep{Fernandez20}. We incorporate this as a two-zone model with a leading `blue' mass with $\kappa_{\rm blue} = 1$\,cm$^2$\,g$^{-1}$ and a deeper layer of `purple' material with $\kappa_{\rm purple} = 5$\,cm$^2$\,g$^{-1}$ \citep[cf.][]{Tanaka20}. This choice reflects the finding that low-$Y_e$ material is ejected the earliest in both hydrodynamic simulations \citep[e.g.][]{Fernandez20} and GRMHD \citep[e.g.][]{Fernandez19b} The fraction of blue mass is determined from an observed relationship with the disc mass via fits to Table 2 of \citet{Fernandez20}. Photons from the purple layer of material must diffuse through the blue layer to reach the observer.

\begin{table}
\begin{adjustbox}{max width=0.475\textwidth,center}
\begin{tabular}{c|cccc}
\hline\hline
Component & Mass & Velocity & Grey opacity & Region \\
 & ($M_{\odot}$) & ($c$) & (cm$^2$\,g$^{-1}$) & (deg) \\
\hline
Dynamical ejecta & KF20 & $0.25$ & 10 & 80 -- 90 \\
Thermal wind & F18, R21 & $0.034$ & 1, 5 & 30 -- 80 \\
Magnetic wind & $m_{\rm therm}$ & $0.22$ & 10 & 0 -- 30 \\
\hline\hline
\end{tabular}
\end{adjustbox}
\caption{Outflow components for our BHNS kilonova model (Gompertz et al. in prep). KF20: \citet{Kruger20}; F18: \citet{Foucart2018}; R21: \citet{Raaijmakers21}. Region angles are from the poles of the spin axis.}
\label{tab:kilonova}
\end{table}

When magnetic fields are included in full three-dimensional general-relativistic magnetohydrodynamic models \citep[e.g.,][]{Siegel17,Siegel18,Fernandez19b} an additional magnetically-driven outflow is identified in polar regions. The dynamics of this ejecta depends on the magnetic field geometry \citep{Christie19}, but it is expected to have a mass roughly equal to the mass of the thermal outflow (i.e. $m_{\rm mag} = m_{\rm therm}$) and an average velocity of $v_{\rm mag} = 0.22$~c \citep{Fernandez19b}. Including this component means that twice the disc fraction $\xi$ derived from the fitting function of \citet{Raaijmakers21} is ejected in winds, in line with expectations from \citet{Fernandez19b} and the \citet{Raaijmakers21} model. The total ejected wind mass is always less than 50 per cent of the accretion disc mass. The magnetic wind is driven from the poles before significant neutrino irradiation can occur, and therefore has $Y_e \sim 0.1$ \citep{Fernandez19b} corresponding to $\kappa_{\rm mag} = 10$\,cm$^2$\,g$^{-1}$. The contribution of the magnetic wind is highly uncertain. Our setup assumes the polar field geometry employed by \citet{Fernandez19b}, but a more toroidal configuration \citep[e.g.][]{Siegel18} should result in a lower ejecta mass and velocity that may make the magnetic wind negligible compared to the thermal wind. Further simulations and/or detections of kilonovae from BHNS mergers will be required to clarify which is the correct picture. See \citet{Christie19} for a discussion on the influence of magnetic fields. Our model is summarised in Table~\ref{tab:kilonova}.

The BHNS ejecta model is integrated as a module in {\sc mosfit} \citep{Guillochon18} which converts the $r$-process masses to light curves through semi-analytical models for the heating rate and deposition \citep{Korobkin12,Barnes16,Cowperthwaite17,Villar17,Metzger19}, and treats the propagation of photons in the common diffusion approximation \citep{Arnett82}. We use the same modules to calculate the photospheric radius \citep{Nicholl17} and the effects of viewing angle \citep{Darbha20} as in \citet{Nicholl21}. The treatment of the kilonova emission is therefore identical to models already published in the literature \citep{Cowperthwaite17,Villar17,Nicholl21}. The masses and velocities of the disc winds and dynamical ejecta that produce this emission are calculated in the same way as other BHNS-driven models for GRBs \citep[e.g.][]{Ascenzi19} and kilonovae \citep[e.g.][]{Barbieri19,Raaijmakers21}, which all follow the scaling relations for material that remains outside of the BH apparent horizon derived by \citet{Kawaguchi16,Foucart2018} and \citet{Kruger20} where available.

\citet{Barbieri19} separate their ejecta into dynamical (grey opacity $\kappa_{\rm dyn} = 15$\,cm$^2$\,g$^{-1}$), wind ($\kappa_{\rm wind} = 1$\,cm$^2$\,g$^{-1}$) and viscous ($\kappa_{\rm vis} = 5$\,cm$^2$\,g$^{-1}$) ejecta. Our model differs due to the inclusion of the magnetically driven wind \citep{Christie19,Fernandez19b} with $\kappa_{\rm mag} = 10$\,cm$^2$\,g$^{-1}$ within 30$^{\circ}$ from the poles. This component is included with a free grey opacity between $1$\,cm$^2$\,g$^{-1} < \kappa_{\rm mag} < 10$\,cm$^2$\,g$^{-1}$ in \citet{Raaijmakers21}. However, their thermal component is assumed to have a much lower grey opacity than ours: $0.1$\,cm$^2$\,g$^{-1} < \kappa_{\rm therm} < 1$\,cm$^2$\,g$^{-1}$. As a result, our model predicts more infra-red/less optical emission for a given set of binary parameters than both \citet{Barbieri19} and \citet{Raaijmakers21}.

\section{Results}
\label{sec:Results}
The ejected mass of the BHNS binary ultimately depends on the zero-age main sequence masses $m_{1,2, \rm ZAMS}$, binary separation $a_{\rm ZAMS}$, and metallicity $Z$, the Maxwellian velocity dispersion $\sigma$ that governs the strength of natal kicks, the breakup spin fraction of the Wolf-Rayet (WR) star $f_{\rm B}$, and the fraction of a donor's envelope that is accreted in stable mass transfer $f_{\rm a}$. In our results, we assume $Z = 0.0002$ which ensures the effect of stellar winds is negligible. Throughout this work, we assume a NS radius of $R_{\rm NS} = 12$ km.

\subsection{Parameter space exploration}
\label{subsec:Param}

\begin{figure}
\centering
\includegraphics[width=0.48\textwidth]{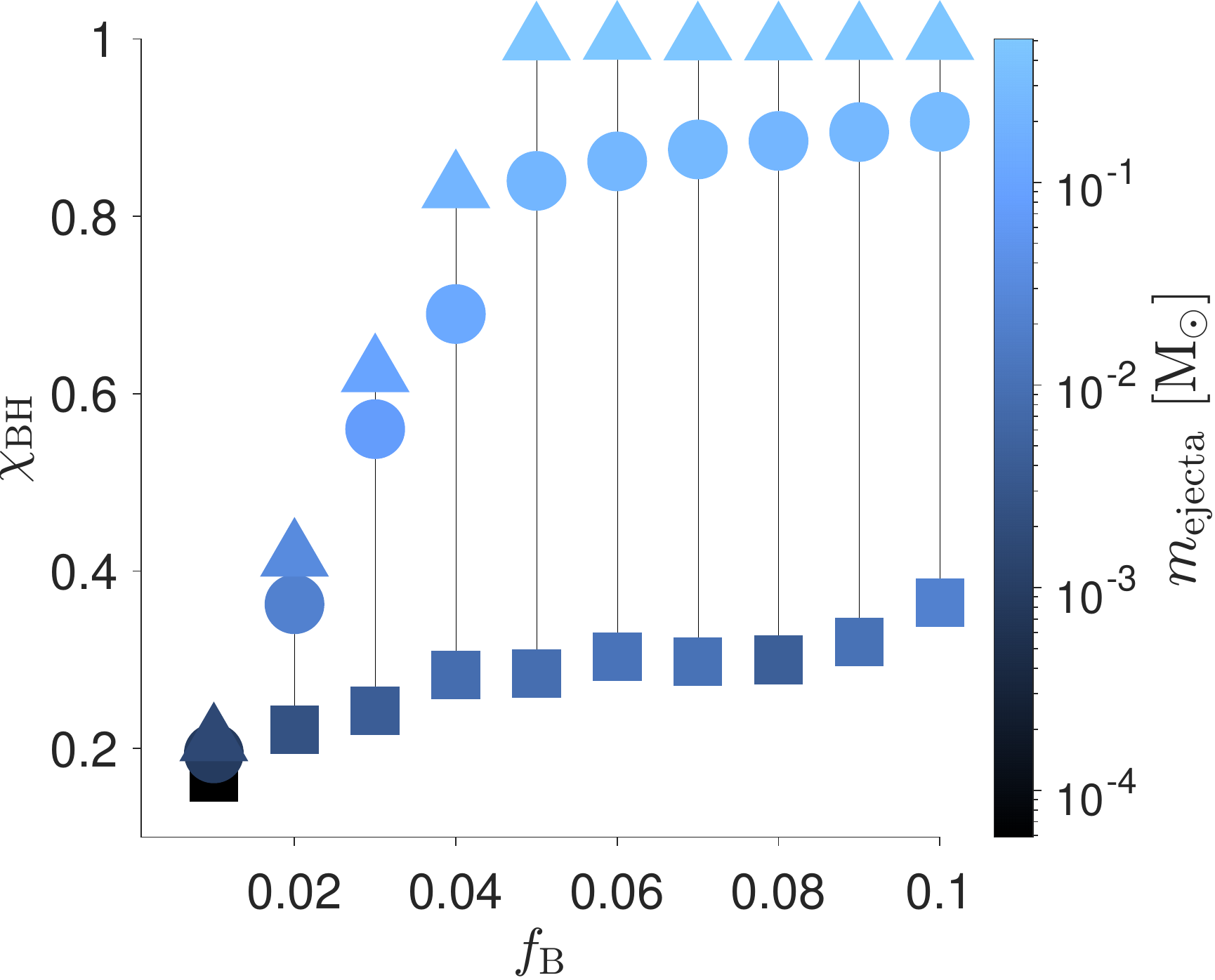}
\caption{
The ejected mass $m_{\rm ejecta}$ of the neutron star and the aligned component of the dimensionless spin of the black hole $\chi_{\rm BH} = \chi_1\cos\theta_1$ for binaries that evolve in Pathway A1 versus the fraction $f_{\rm B}$ of the breakup spin with which the Wolf-Rayet progenitor star of the black hole is born. The stellar binaries are initialized with $m_{1,\rm ZAMS} = 20$ M$_{\odot}$, $m_{2,\rm ZAMS} = 13$ M$_{\odot}$, $a_{\rm ZAMS} = 6,000$ R$_{\odot}$, $f_{\rm a} = 0$, and $\sigma = 30$ km/s, and assuming weak core-envelope coupling and negligible mass loss in formation due to the Kerr limit. The shape of the markers, connected by a vertical line for each value of $f_{\rm B}$, represent the 95$^{\rm th}$ (triangle), 50$^{\rm th}$ (circle), and 5$^{\rm th}$ (square) percentiles of $\chi_{\rm BH}$, and the color of the markers indicates the corresponding percentiles of the ejected mass $m_{\rm ejecta}$.
} \label{F:PathwayA1}
\end{figure}

\begin{figure*}
\centering
\includegraphics[width=\textwidth]{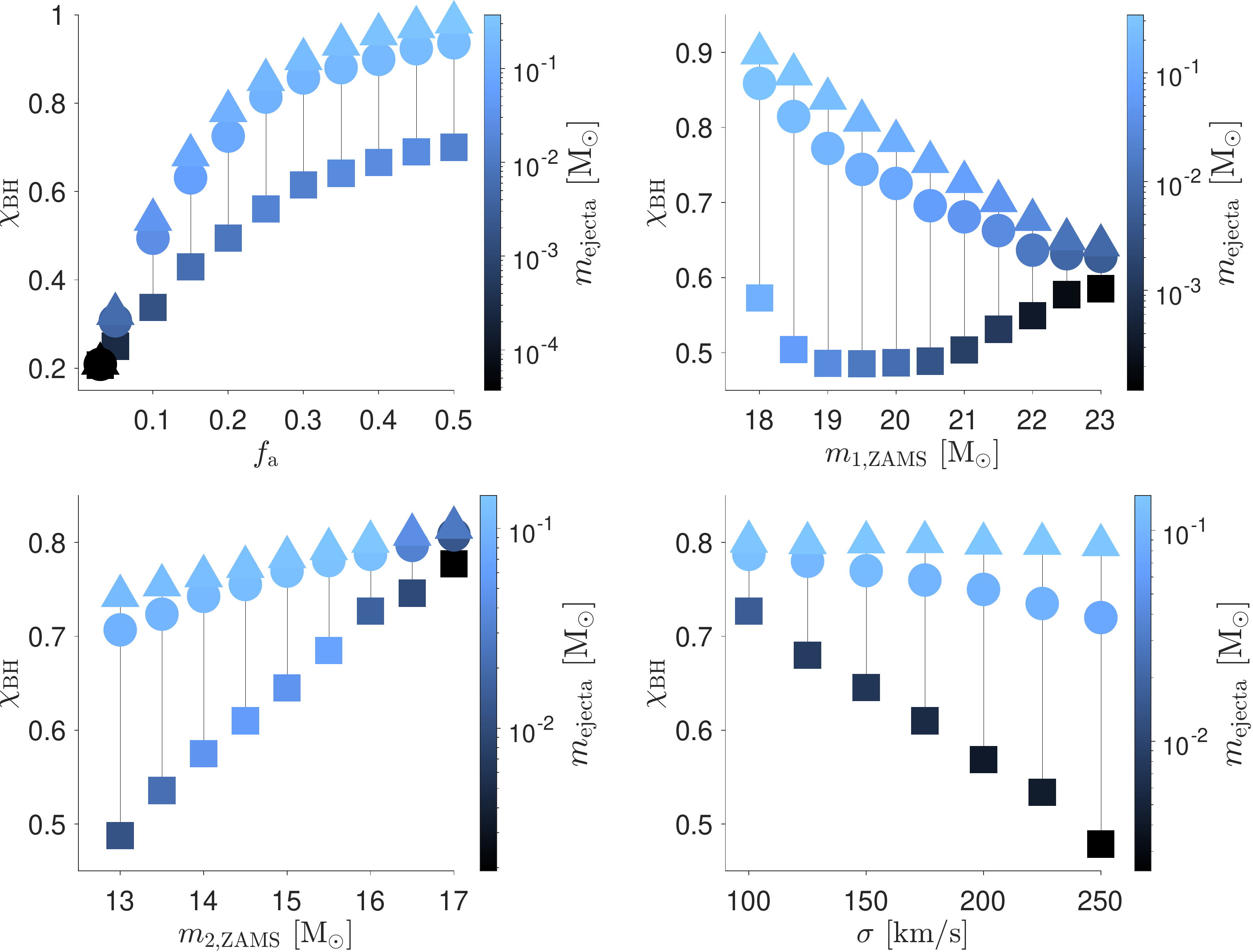}
\caption{The ejected mass $m_{\rm ejecta}$ of the neutron star and the aligned component of the dimensionless spin of the black hole $\chi_{\rm BH} = \chi_1\cos\theta_1$ as functions of the fraction $f_{\rm a}$ that is accreted in stable mass transfer by the black hole (top-left panel), the initial mass $m_{1,\rm ZAMS}$ of the primary star (top-right panel), the initial mass $m_{2,\rm ZAMS}$ of the secondary star (bottom-left panel), and the Maxwellian velocity dispersion $\sigma$ of the natal kicks (bottom-right panel). The stellar binaries in the top-left panel are initialized with $m_{1,\rm ZAMS} = 20$ M$_{\odot}$, $m_{2,\rm ZAMS} = 15$ M$_{\odot}$, $a_{\rm ZAMS} = 12{,}000$ R$_{\odot}$, and $\sigma = 200$ km/s, in the top-right panel with $f_{\rm a} = 0.2$, $m_{2,\rm ZAMS} = 15$ M$_{\odot}$, $a_{\rm ZAMS} = 12{,}000$ R$_{\odot}$, and $\sigma = 200$ km/s, in the bottom-left panel with $f_{\rm a} = 0.2$, $m_{1,\rm ZAMS} = 20$ M$_{\odot}$, $a_{\rm ZAMS} = 12{,}000$ R$_{\odot}$, and $\sigma = 200$ km/s, and in the bottom-right panel with $f_{\rm a} = 0.2$, $m_{1,\rm ZAMS} = 20$ M$_{\odot}$, $m_{2,\rm ZAMS} = 15$ M$_{\odot}$, and $a_{\rm ZAMS} = 12{,}000$ R$_{\odot}$. In each panel we assume strong stellar core-envelope coupling and isotropic mass loss in formation due to the Kerr limit. The shape of the markers, connected by a vertical line for each value of $f_{\rm B}$, represents the 95$^{\rm th}$ (triangle), 50$^{\rm th}$ (circle), and 5$^{\rm th}$ (square) percentiles of $\chi_{\rm BH}$, and the color of the markers indicates the corresponding percentiles of the ejected mass $m_{\rm ejecta}$.
} \label{F:PathwayB1_all}
\end{figure*}

The binaries presented in Figures \ref{F:PathwayA1} and \ref{F:PathwayB1_all} are distributions of BHNSs where one free parameter is varied (the horizontal axis) and all others are held constant. These figures depict the 5$^{\rm th}$, 50$^{\rm th}$, and 95$^{\rm th}$ percentiles of the dimensionless spin of the BH $\chi_{\rm BH} = \chi_1\cos\theta_1$ (the vertical axis) and the corresponding percentiles of the ejected mass $m_{\rm ejecta}$ of the NS with a colorbar. For fixed binary mass ratio and NS compactness, $m_{\rm ejecta}$ is monotonic in $\chi_{\rm BH}$.
As we evolve the spin magnitudes and directions until BHNS formation, $\chi_{\rm BH}$ depends principally on the mechanism by which the BH acquires spin (i.e., inheritance or accretion) and on the natal kick velocity dispersion $\sigma$. These isotropically oriented natal kicks produce scatter in the BH misalignments $\cos\theta_1$ that is preferentially peaked near $\cos\theta_1 \approx 1$ since the ZAMS spins are assumed to be aligned with the binary orbital angular momentum. 

The first mechanism that we explore to obtain a highly spinning BH is inheritance via weak core-envelope coupling for binaries that evolve in Pathway A1. The WR breakup spin fraction $f_{\rm B}$ determines $\chi_{\rm BH}$ and $m_{\rm ejecta}$ as shown in Fig.~\ref{F:PathwayA1}. In the limit of small inherited spins, i.e., $f_{\rm B} < 0.01$, the aligned BH spin is also very small causing the NS to be captured rather than tidally disrupted, and hence $m_{\rm ejecta} = 0$ by definition. As the BH inherits larger spin, $f_{\rm B} \gtrsim 0.01$, the NS can be tidally disrupted allowing for nonzero $m_{\rm ejecta}$ and hence an observable counterpart. The inherited spin of the BH increases linearly with larger $f_{\rm B}$, as $\chi_1 \propto f_{\rm B}\chi_{\rm B}$ and $m_{\rm 1,ZAMS}$ is held constant (see Eq.~(\ref{E:Break})). Meanwhile, the scatter in the aligned BH spin component $\chi_{\rm BH}$ increases with $f_{\rm B}$ because the distribution of misalignments is biased towards $\cos\theta_1 \sim 1$ with a tail to larger values for this constant value of $\sigma$. For $f_{\rm B} \sim 0.03$, $\chi_{\rm BH}$ becomes sufficiently large to yield significant ejected mass, i.e., $m_{\rm ejecta} \gtrsim 0.01$ M$_{\odot}$.
For $f_{\rm B} \gtrsim 0.05$, the 95$^{\rm th}$ percentile (triangles) of $m_{\rm ejecta}$ is largest as the spin magnitude of the BH $\chi_1$ is maximal implying that $\chi_{\rm BH}$ asymptotes at 1, and $\chi_{\rm BH}$ corresponding to the median (circles) of $m_{\rm ejecta}$ approaches $\approx 0.9$. The masses of the BH and of the NS are the same for each value of $f_{\rm B}$, and if either were larger then $m_{\rm ejecta}$ would be suppressed. The kick velocity dispersion which provides the scatter in $\chi_{\rm BH}$ is $\sigma = 30$ km/s for these binaries, implying that larger $\sigma$ could decrease the median of $m_{\rm ejecta}$. The BH and NS masses are fixed in these distributions with the mass ratio $Q = m_{\rm BH}/m_{\rm NS} \simeq 3.5$.

The second mechanism we explore is accretion during stable mass transfer for binaries that evolve in Pathway B1 with the assumption of strong core-envelope coupling. Fig.~\ref{F:PathwayB1_all} displays the dependence of $\chi_{\rm BH}$ and $m_{\rm ejecta}$ on the accreted fraction $f_{\rm a}$ (top-left panel), the initial mass of the primary star $m_{1, \rm ZAMS}$ (top-right panel), the initial mass of the secondary star $m_{2, \rm ZAMS}$ (bottom-left panel), and the natal kick strength $\sigma$ (bottom-right panel). As the primary undergoes core collapse prior to the loss of the envelope of the secondary star, the primary accretes as a BH. 
In the top-left and top-right panels, $m_{\rm BH}$ ranges from $\simeq$ 5 to 9~M$_{\odot}$ as $f_{\rm a}$ and $m_{1,\rm ZAMS}$ increase, respectively, and $m_{\rm NS} \simeq 1.3$~M$_{\odot}$ is constant. In the bottom-left panel $m_{\rm BH}$ ranges from $\simeq 6.5$ to 7~M$_{\odot}$ and $m_{\rm NS}$ increases sharply from $\simeq 1.3$ to 1.7~M$_{\odot}$ at $m_{2,\rm ZAMS} = 16$~M$_{\odot}$ due to the rapid supernova mechanism employed from \citet{Fryer2012}. In the bottom-right panel the masses $m_{\rm BH} \simeq 6.7$~M$_{\odot}$ and $m_{\rm NS} \simeq 1.3$~M$_{\odot}$ are constant.

In the top-left panel, the values of $m_{1, \rm ZAMS}$, $m_{2, \rm ZAMS}$, and $\sigma$ are fixed while $f_{\rm a}$ is varied. The BH spin magnitude $\chi_1$ remains small for a small amount of accretion $f_{\rm a} \lesssim 0.1$ due to dissipation of the stellar progenitor's spin via strong core-envelope coupling during common envelope evolution. The resultant ejected mass is small. As $f_{\rm a}$ increases, the BH accretes more gas from its companion's envelope resulting in a larger spin magnitude $\chi_1$ and ejected mass $m_{\rm ejecta}$. However, the increase of $\chi_1$ is nonlinear in $f_{\rm a}$ because the energy $E(\chi)$ of an accreted particle (Eq.~(\ref{E:BHAcc1})) is roughly constant with an increasing amount of accreted mass while the orbital angular momentum $L(\chi)$ of an accreting particle decreases due the smaller $R_{\rm ISCO}$ that results from the higher BH spin. In the limit of large accretion $f_{\rm a} \gtrsim 0.4$, the BH spin is high $\chi_1 > 0.5$ and increases only slightly. The ejected mass is significant for even the 5$^{\rm th}$ percentile of binaries, suggesting that accretion onto the primary BH may be a promising mechanism for producing observable counterparts. 

\begin{figure*}
\centering
\includegraphics[width=\textwidth]{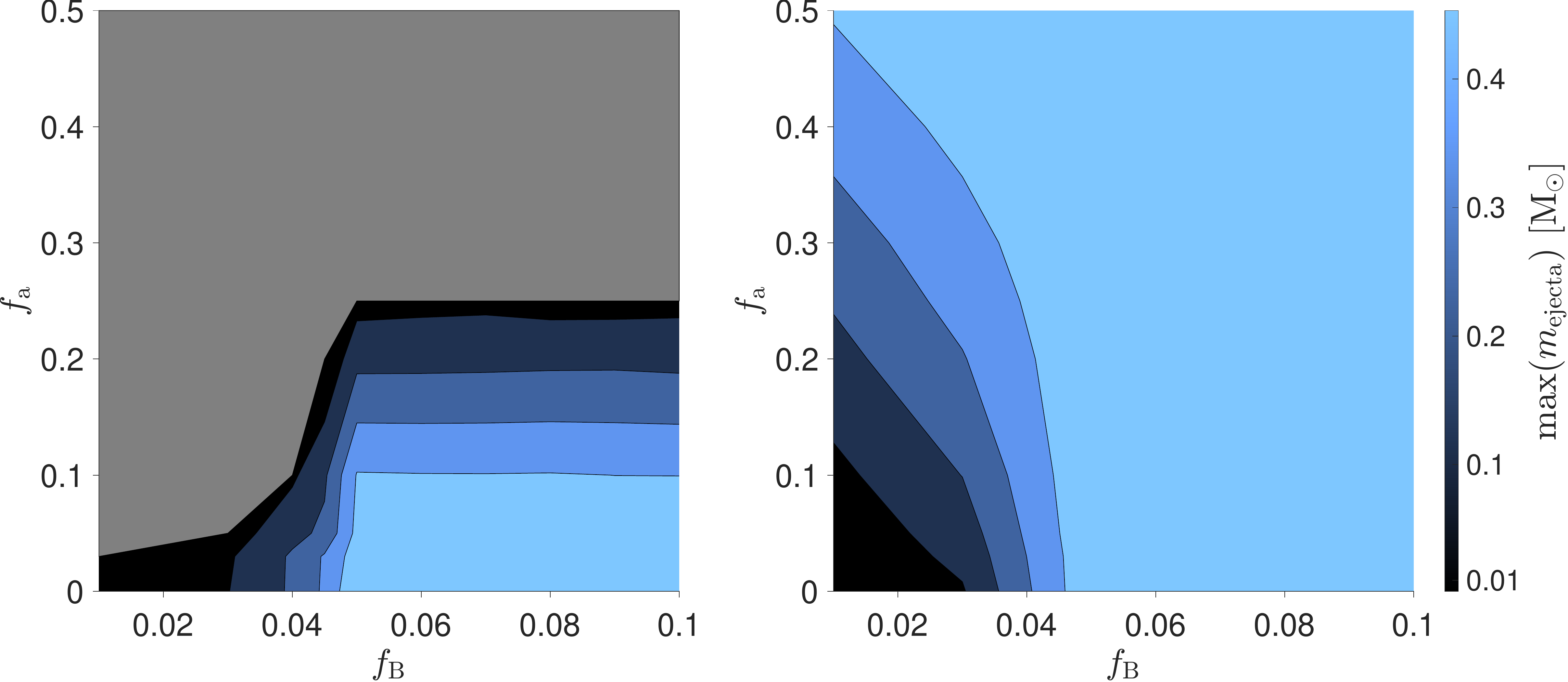}
\caption{Contours of the maximum, i.e., the 100$^{\rm th}$ percentile, of the ejected mass $m_{\rm ejecta}$ as functions of the fraction $f_{\rm a}$ of the donor's envelope that is accreted in stable mass transfer by the black hole and of the fraction $f_{\rm B}$ of the breakup spin with which the Wolf-Rayet progenitor star of the black hole is born. In the left panel are binaries that evolve from Pathway A1, where the secondary star accretes as a main sequence star, and in the right panel are binaries that evolve from Pathway B1, where the primary star accretes after forming a black hole. The stellar binaries in the left panel (right panel) are initialized with $m_{1,\rm ZAMS} = 20$ M$_{\odot}$, $m_{2,\rm ZAMS} = 15$ M$_{\odot}$, $\sigma = 30$ km/s ($\sigma = 200$ km/s), and $a_{\rm ZAMS} = 6{,}000$ R$_{\odot}$ ($a_{\rm ZAMS} = 12{,}000$ R$_{\odot}$), and assuming weak core-envelope coupling and negligible (isotropic) mass loss in formation due to the Kerr limit. The grey region in the left panel represents binaries whose neutron star is not tidally disrupted.
} \label{F:ContoursFaFb}
\end{figure*}

In the top-right panel of Fig.~\ref{F:PathwayB1_all} the values of $f_{\rm a}$, $m_{2, \rm ZAMS}$, and $\sigma$ are fixed while $m_{1, \rm ZAMS}$ is varied, i.e., the initial mass of the BH accretor increases with increasing $m_{1, \rm ZAMS}$. Since the amount of gas that is accreted is held constant here, the spin of the BH generally decreases as the initial mass of the BH increases, as seen in Eq.~(\ref{E:BHAcc2}) where $m_{\rm BH}$ is in the denominator. Simultaneously, the radius $R_{\rm ISCO}$ of the BH is larger for smaller $m_{1, \rm ZAMS}$ due to the small BH mass but also because the BH spin is high, implying that $m_{\rm ejecta}$ is at its largest. The scatter in the aligned BH spin $\chi_{\rm BH}$ generally decreases for increasing $m_{1, \rm ZAMS}$ as the natal kick velocity is suppressed from fallback accretion and a larger orbital velocity. For $m_{1, \rm ZAMS} \gtrsim 22.5$ M$_{\odot}$, fallback completely suppresses the natal kick of the primary, and the misalignment solely originates from the natal kick of the secondary. In the limit of large $m_{1, \rm ZAMS}$, the BH spin is $\chi_1 \sim 0.6$, consistent with the results of \citet{Steinle2021}. 

The dependence on $m_{2,\rm ZAMS}$, shown in the bottom-left panel of Fig.~\ref{F:PathwayB1_all} where $f_{\rm a}$, $m_{1,\rm ZAMS}$, and $\sigma$ are fixed, is complicated from the interplay of competing factors. For $m_{2,\rm ZAMS} < 16$ M$_{\odot}$, the mass of the NS is constant. The spin of the accreting BH increases with increasing $m_{2,\rm ZAMS}$ as the envelope of the secondary, and hence the amount of gas to be accreted, increases, implying a larger $m_{\rm ejecta}$. The scatter in $\chi_{\rm BH}$ decreases with increasing $m_{2,\rm ZAMS}$ which increases the orbital velocity prior to the second SN. For $m_{2,\rm ZAMS} \geq 16$ M$_{\odot}$, the mass of the NS at formation is larger increasing the NS compactness and decreasing $m_{\rm ejecta}$ despite the larger $\chi_{\rm BH}$. This competition between the increase in the BH spin from a larger donor's envelope in stable mass transfer and the increase in the NS compactness is a distinct feature, but is possibly model dependent since the dependence of the mass of the NS on $m_{2,\rm ZAMS}$ in uncertain.

The dependence on $\sigma$ is shown in the bottom-right panel of Fig.~\ref{F:PathwayB1_all} with fixed $f_{\rm a}$, $m_{1,\rm ZAMS}$, and $m_{2,\rm ZAMS}$. Larger values of $\sigma$ generally produce larger spin-orbit misalignments and smaller $\chi_{\rm BH}$, which suppresses $m_{\rm ejecta}$. The 5$^{\rm th}$ percentile of $\chi_{\rm BH}$ remains roughly constant with increasing $\sigma$ because the distributions of $\cos\theta_1$ are biased toward unity. This dependence is similar for binaries that evolve in A1, except for smaller values of $\sigma$ since common envelope evolution occurs prior to the natal kick of the primary.

Together, these results demonstrate that although we can produce rapidly spinning BHs in the isolated formation channel through the mechanisms of inheritance or accretion, the uncertainties of stellar binary evolution also affect other parameters such as spin-orbit misalignments and the binary mass ratio, which themselves affect the resultant ejected mass. Therefore, a question naturally arises: how would one distinguish between the two possible formation pathways explored here? Answering this question for a single (population of) observed BHNS(s) with statistical rigor would require (hierarchical) Bayesian parameter estimation. Although such an analysis is beyond the scope of this work, we can identify regions of the parameter space that are likely to favor systems with observable electromagnetic counterparts from either pathway.

\begin{figure*}
\centering
\includegraphics[width=\textwidth]{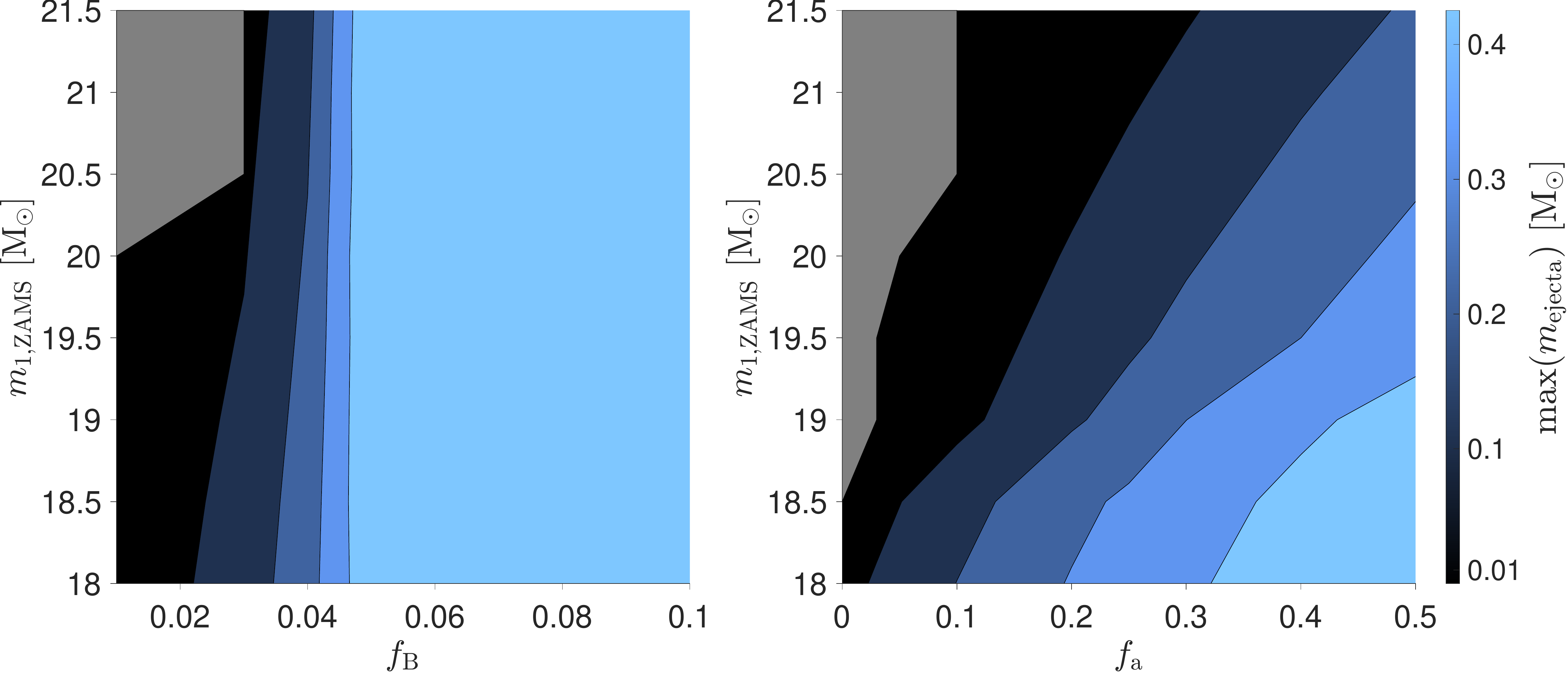}
\caption{Contours of the maximum, i.e., the 100$^{\rm th}$ percentile, of the ejected mass $m_{\rm ejecta}$ as functions of the ZAMS mass of the primary $m_{1, \rm ZAMS}$, which forms the black hole, and of a mechanism for obtaining high black spin spin. For binaries that evolve in Pathway A1 (left panel), we consider the fraction $f_{\rm B}$ ($f_{\rm a}$) of the breakup spin with which the Wolf-Rayet progenitor star of the black hole is born under the assumption of weak stellar core-envelope coupling. For binaries that evolve in Pathway B1 (right panel), we consider the fraction of the donor's envelope that is accreted in stable mass transfer by the black hole under the assumption of strong core-envelope coupling. The stellar binaries in the left panel (right panel) are initialized with $m_{2,\rm ZAMS} = 13$ M$_{\odot}$, $\sigma = 30$ km/s ($\sigma = 200$ km/s), and $a_{\rm ZAMS} = 6{,}000$ R$_{\odot}$ ($a_{\rm ZAMS} = 12{,}000$ R$_{\odot}$), and assuming negligible (isotropic) mass loss in formation due to the Kerr limit. The grey regions represent binaries whose neutron star is not tidally disrupted.
} \label{F:ContoursMassFracs}
\end{figure*}

The two panels of Figure~\ref{F:ContoursFaFb} depict contours of the maximum possible ejected mass, i.e., the 100$^{\rm th}$ percentile of $m_{\rm ejecta}$, under the assumption of weak core-envelope coupling. For binaries that evolve in Pathway A1, shown in the left panel (where $m_{\rm BH} = 4.6$~M$_{\odot}$ is constant and $m_{\rm NS}$ ranges from $\simeq 1.3$ to 2.5~M$_{\odot}$ over $f_{\rm a}$), accretion is not an efficient mechanism for producing significant ejected mass, as the secondary star accretes during stable mass transfer of the primary. Indeed, a small amount of accretion, i.e. $f_{\rm a} \lesssim 0.1$, can result in large $m_{\rm ejecta}$ if the BH inherits a high natal spin, i.e., $f_{\rm B} \gtrsim 0.05$, because the mass of the NS is not too large. Increased accretion onto the secondary main sequence star cause the mass (and hence compactness) of the NS that forms to be subsequently too large which suppresses $m_{\rm ejecta}$. For $f_{\rm a} \sim 0.25$, the NS is not tidally disrupted, indicated by the grey region. This is even more prominent for sufficiently small BH natal spins $f_{\rm B} \lesssim 0.05$, as a smaller NS mass can prevent tidal disruption for smaller BH spin. In the limit of no accretion $f_{\rm a} \sim 0$ and negligible inherited BH spin $f_{\rm B} \lesssim 0.01$, the maximum ejected mass $m_{\rm ejecta} \approx 0.001$, consistent with the 95$^{\rm th}$ percentile in Fig.~\ref{F:PathwayA1}. If we instead assume that mass loss in BH formation due to the Kerr limit is isotropic rather than negligible, then more accreted mass can yield larger $m_{\rm ejecta}$ due to the primary BH mass being smaller, although this effect is not significant even for $f_{\rm B} \gtrsim 0.05$.

Binaries that evolve in Pathway B1 are shown in the right panel of Fig.~\ref{F:ContoursFaFb}, where the the primary accretes as a BH during stable mass transfer of the secondary. Here $m_{\rm BH}$ ranges from $\simeq 5$ to 9~M$_{\odot}$ over $f_{\rm a}$, and $m_{\rm NS} \simeq 1.3$~M$_{\odot}$ is constant. For small inherited natal spin and small amounts of accretion, the maximum $m_{\rm ejecta}$ is small due to the small BH spin. As either $f_{\rm a}$ or $f_{\rm B}$ are increased, the BH spin increases and allows for a larger maximum $m_{\rm ejecta}$. Consistent with the 95$^{\rm th}$ percentile in the left panel of Fig.~\ref{F:PathwayB1_all}, an accreted fraction $f_{\rm a} \sim 0.5$ can produce high BH natal spin and $m_{\rm ejecta}$ as large as that from large inherited spin, i.e., $f_{\rm B} \sim 0.05$ in in Fig.~\ref{F:PathwayA1}.

The left (right) panel of Figure~\ref{F:ContoursMassFracs} displays contours of the maximum of $m_{\rm ejecta}$ as functions of $m_{1,\rm ZAMS}$ and $f_{\rm B}$ ($f_{\rm a}$) for binaries that evolve in Pathway A1 (B1) under the assumption of weak (strong) core envelope coupling. In the left (right) panel for A1 (B1), $m_{\rm BH}$ ranges from $\simeq 3.1$ to 6~M$_{\odot}$ over $m_{1,\rm ZAMS}$ (from $\simeq 3.1$ to 10~M$_{\odot}$ over $f_{\rm a}$ and $m_{1,\rm ZAMS}$) and $m_{\rm NS} \simeq 1.3$~M$_{\odot}$ is constant.
In both pathways, the mass of the BH increases as $m_{1,\rm ZAMS}$ increases, providing a larger mass ratio $Q$ which suppresses $m_{\rm ejecta}$. In the limit of small BH spins, i.e., $f_{\rm B} \to 0$ in the left panel and $f_{\rm a} \to 0$ in the right panel, sufficiently large $m_{1,\rm ZAMS}$ disallows tidal disruption of the NS, indicated by the grey region. In Pathway A1, $m_{\rm ejecta}$ increases sharply as $f_{\rm B}$ increases, consistent with Fig.~\ref{F:PathwayA1}, and reaches a maximum of $m_{\rm ejecta} \sim 0.4$ M$_{\odot}$ for $f_{\rm B} \gtrsim 0.05$ due to maximal BH spin. 
Comparatively, in Pathway B1 $m_{\rm ejecta}$ increases gradually as $f_{\rm a}$ increases from 0, because the mass of the BH increases from gas accretion. This implies that smaller $m_{1,\rm ZAMS}$ is needed in Pathway B1 than in A1 to obtain very large $m_{\rm ejecta}$. However, if accretion in B1 is highly conservative, i.e., $f_{\rm a} \sim 0.9$, the spin of the BH is near maximal allowing for larger $m_{\rm ejecta}$ at higher values of $m_{1,\rm ZAMS}$.

Comparing the contours in Fig.~\ref{F:ContoursFaFb} between binaries that evolve from these two pathways, the asymmetry in the effect of accretion allows NSs from Pathway B1 (right panel) to be tidally disrupted and produce significant ejected mass essentially over the entire spin parameter space, whereas NSs from Pathway A1 (left panel) are not tidally disrupted over half of this region of the spin parameter space. On the other hand, comparing the contours in Fig.~\ref{F:ContoursMassFracs}, the mass of the BH can suppress the ejected mass in B1 more than in A1 due to increased BH mass from non-conservative accretion. This asymmetry provides signatures to distinguish these two possible formation pathways if the values of $f_{\rm a}$, $f_{\rm B}$, and $m_{1,\rm ZAMS}$ can be constrained for a population of BHNS binaries observed from gravitational-wave data, though $m_{\rm BH}$ as measured from gravitational-wave detections is degenerate with $f_{\rm a}$ and $m_{1,\rm ZAMS}$. Additionally, a systematically larger mass ratio $Q = m_{\rm BH}/m_{\rm NS}$ could be expected for binaries that evolve from Pathway B1 than in A1 due to accretion by the BH in Pathway B1.

If the ejected masses can be measured from electromagnetic follow-up for a population of BHNS binaries whose BH spins are measured from gravitational-wave data, hierarchical Bayesian parameter estimation could constrain the likely source of the spin of the BH. In such an inference study, one could create an astrophysical model by leveraging the fact that there would be a stronger correlation between the BH mass and $f_{\rm a}$ than between the BH mass and $f_{\rm B}$, as shown by the contours of Fig.~\ref{F:ContoursMassFracs}. Naively, one could expect these correlations to be opposite, however this is highly model dependent as the relationship between $m_{1,\rm ZAMS}$ and $f_{\rm B}$ depends on the strength of core-envelope coupling, of which we simply consider extreme limits, and the mechanisms that drive stellar angular momentum transport, which are uncertain. Subsequently placing the constraints for the population of binaries in the planes of Fig.'s~\ref{F:ContoursFaFb} and \ref{F:ContoursMassFracs} could therefore elucidate possible sources of BH spin and ejected mass, and possible formation pathways. 
Although we show the maximum possible ejected mass in Fig.'s~\ref{F:ContoursFaFb} and \ref{F:ContoursMassFracs}, the results do not change qualitatively if we instead use the median of $m_{\rm ejecta}$, implying that events with somewhat faint kilonova counterparts can be understood within these trends. However, these results are best interpreted in the context of bright kilonova.

\subsection{Light curves of electromagnetic counterparts}\label{subsec:lightcurves}

\begin{figure}
    \centering
    \includegraphics[width=\columnwidth]{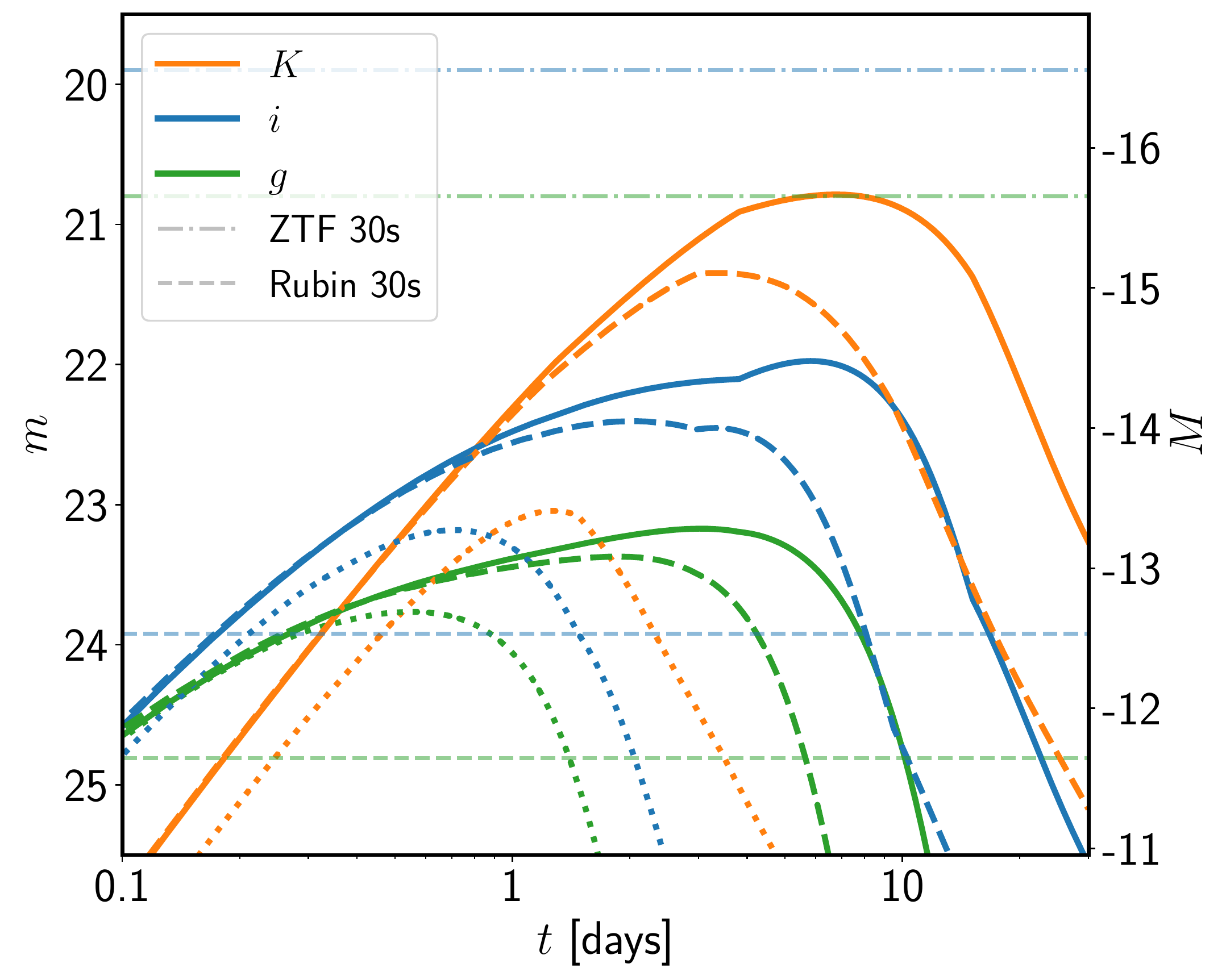}
    \caption{Predicted kilonova light curves from the model described in Section~\ref{subsec:lightcurves} at an assumed distance of 200\,Mpc. The orange, blue, and green lines correspond to the $K$-band, $i$-band, and $g$-band light curves, respectively, and each band is shown for the three orbit-aligned BH spins $\chi_{\rm BH} = 0.2$, $0.56$ and $0.9$ that correspond to the dotted, dashed, and solid curves, respectively. The $5\sigma$ median $g$- and $i$-band limiting magnitudes from a 30s exposure with \emph{Zwicky} Transient Facility \citep[$m_g > 20.8$; $m_i > 19.9$;][]{Bellm19} and the Vera Rubin Observatory \citep[$m_g > 24.81$; $m_i > 23.92$;][]{Ivezic19} are shown by the horizontal dot-dash and dashed lines, respectively.
    }
    \label{fig:lightcurves}
\end{figure}

We consider three values of the aligned BH spin component $\chi_{\rm BH} =$ 0.2, 0.56, and 0.9 which correspond to the medians of $\chi_{\rm BH}$ for the BHNSs from Section~\ref{subsec:Param} that evolved under weak core-envelope coupling with $f_{\rm B} =$ 0.01, 0.03, and 0.1, respectively (i.e., see Fig.~\ref{F:PathwayA1}). These BHNSs have the same mass ratio $Q = m_{\rm BH}/m_{\rm NS} = 4.7/1.3 = 3.6$ and $R_{\rm NS} = 12$ km, and yield $m_{\rm ejecta} \sim 10^{-3},~10^{-2}$, and $10^{-1}$ M$_{\odot}$, respectively. Though we did not show this in Subsection~\ref{subsec:Param}, note that higher $Q$ or a more compact NS can result in a lower ejecta mass from Eq.~(\ref{E:EjectedMass}) and therefore a fainter electromagnetic transient. We assume the observer is oriented 30$^{\circ}$ from the pole, at the boundary between the thermal and magnetic disc wind outflows. This is done in order to sample all three emission components and obtain an `average' kilonova, though in practice it is an arbitrary choice. For $Q = 3.6$, $v_{\rm dyn} = 0.25c$ \citep{Kawaguchi16}.

The resultant light curves for a merger at an assumed distance of 200\,Mpc are shown in Figure~\ref{fig:lightcurves}. Their morphology is driven by the interplay of the three emission components (Table~\ref{tab:kilonova}) whose relative contributions depend on the properties of the input binary \citep[cf.][]{Kawaguchi16,Foucart2018,Kruger20,Raaijmakers21}. Each component contributes radiation at different temperatures and on different timescales, resulting in time- and frequency-dependent light curve behaviour \citep[e.g.][]{Metzger10,Grossman14,Kasen15}. The fraction of Lanthanides and Actinides in the ejecta is particularly impactful in this regard; the complex absorption patterns of these more massive elements absorb much of the light at optical frequencies, resulting in `red' (infra-red bright) emission \citep{Barnes13,Kasen13,Tanaka13,Kasen15,Tanaka20}. This can be broadly understood from the approximated grey opacities in Table~\ref{tab:kilonova}, where emission components with higher values contribute more in the infra-red ($K$-band, orange lines) at later times, while those with lower values produce optical emission ($g$-band, green lines) earlier in the evolution. The $i$-band (blue lines) is intermediate in frequency between the two. The total ejecta mass depends strongly on $\chi_{\rm BH}$ \citep[e.g.][]{Etienne2009,Kyutoku2011,Lovelace13,Kyutoku15,Foucart2018}, hence higher spin models result in more massive winds and dynamical outflows that produce brighter and longer-lived emission.

Our model shows that while kilonovae from BHNS mergers at this distance are likely not detectable by the current generation of survey telescopes like the \emph{Zwicky} Transient Facility \citep[ZTF;][]{Bellm19}, the Asteroid Terrestrial Impact Last Alert System \citep[ATLAS;][]{Tonry18} or the Gravitational-wave Optical Transient Observer \citep[GOTO;][]{Dyer20,Steeghs22}, these events are expected to be detectable by the Vera Rubin Observatory \citep{Ivezic19,Andreoni22}. Such a finding is in line with the non-detections of BHNS merger candidates during O3 \citep{Hosseinzadeh19,Lundquist19,Ackley20,Andreoni20,Antier20,Gompertz20,Anand21,Oates21,Paterson21}, though GW-triggered events are likely to be probed to greater depths than untriggered survey observations.

In addition to the kilonova described above, we can estimate the power of any short GRB that is launched. For aligned spins of $\chi_{\rm BH} = 0.2$, $0.56$ and $0.9$, we estimate that $2.0\times10^{-3}$\,$M_{\odot}$, $3.0\times10^{-2}$\,$M_{\odot}$, and $1.4\times10^{-1}$\,$M_{\odot}$ accretes onto the remnant black hole after accounting for disc wind outflows. The luminosity of the resultant jet can be approximated by $L = \epsilon M_{\rm acc}/t_{\rm acc}c^2$\,erg\,s$^{-1}$, where $M_{\rm acc}$ is the mass accreted in time $t_{\rm acc}$ and $\epsilon$ is the efficiency in converting accretion power to electromagnetic luminosity \citep[cf.][]{Ruiz21}. For efficiencies of the order of 1 per cent and accretion timescales $\lesssim 2$s our accreted masses provide reasonable agreement with the observed short GRB population, which have typical luminosities in the 15--150\,keV \emph{Swift}-BAT bandpass of $\sim 10^{49}$\,erg\,s$^{-1}$ \citep[e.g.][assuming a $10^{\circ}$ jet opening angle]{Gompertz20b}. However, the most massive discs, particularly those corresponding to $\chi_{\rm BH} = 0.9$, require accretion timescales of the order of a hundred seconds to produce suitable luminosities. This is consistent with the population of `extended emission' (EE) GRBs \citep{Norris06,Norris10,Gompertz13} which were recently linked to compact object mergers through the detection of a kilonova alongside GRB\,211211A \citep[][though see \citealt{Waxman22} for an alternative interpretation]{Gompertz22b,Rastinejad22,Troja22} and have been suggested to arise from BHNS mergers \citep{Troja08,Gompertz20b}. Such long accretion timescales may be achieved through late fall-back from marginally bound tidal tails \citep{Rosswog07,Desai19}.

\section{Conclusions and Discussion}
\label{sec:Disc}
The possibility of observing electromagnetic counterparts, i.e., short GRBs and kilonovae, from the merger of BHNS binaries is an exciting prospect for multi-messenger astronomy. The existence and detectability of these counterparts is sensitive to the properties of the BHNSs that produce them, implying that accurate modeling of populations of BHNS binaries is crucial for understanding the prevalence of counterparts in the Universe. Currently, there are great uncertainties in models of BHNS binary formation and spin evolution.

The most important BHNS properties for producing a large amount of ejecta mass $m_{\rm ejecta}$, and hence a detectable counterpart, are the masses of the BH and NS, and the aligned component of the BH spin $\chi_{\rm BH}$. We explored the dependence of these quantities on two key mechanisms by which BHs in BHNS binaries may obtain high spin magnitude. Either the BH inherits spin from its Wolf-Rayet stellar progenitor that evolved under weak core-envelope coupling with a fraction $f_{\rm B}$ of its maximum breakup spin, or the BH gains spin by accreting a fraction $f_{\rm a}$ of its companion's envelope.
Significant $m_{\rm ejecta}$ is possible from:
\begin{enumerate}[leftmargin=*]
    \item Inheritance of high BH spin $\chi_1$ via weak core-envelope coupling with $f_{\rm B} \gtrsim 0.03$, where $\chi_{\rm BH}$ scales linearly with $f_{\rm B}$ until the Kerr limit is saturated.
    
    \item Accretion of high BH spin, where the BH spin scales nonlinearly with $f_{\rm a}$. The mass of the BH increases with increasing $f_{\rm a}$, which suppresses $m_{\rm ejecta}$ if the initial mass of the BH is too large.
    
    \item Spin-orbit misalignments that are not too large, where the source of misalignments are natal kicks.
\end{enumerate}
Instead, higher $Z$ would produce a smaller BH spin magnitude due to stellar winds that suppresses the amount of ejected mass and the brightness of a possible kilonova, but could also produce a lower BH mass that enhances the ejected mass.

We also identify signatures of $f_{\rm B}$ and $f_{\rm a}$ on the BH spin and possible ejected mass in two formation pathways defined as A1 (B1) when the primary star undergoes stable mass transfer (common envelope evolution) and the secondary undergoes common envelope evolution (stable mass transfer). 
An asymmetry exists between the ejected mass of BHNSs in these pathways, as moderate, not highly-conservative accretion can provide a high BH spin and significant ejected mass in B1 but not in A1. For a given stellar initial mass function we predict more high-ejecta-mass kilonovae for binaries that evolve in Pathway A1 if typically $f_{\rm B} \gtrsim 0.04$, while kilonovae from binaries that evolve in Pathway B1 can be limited by increased BH mass from (moderate, not highly-conservative) accretion. We assumed a low metallicity $Z = $ \num{2e-4}. 

As the simple model we consider here is not capable of producing realistic predictions of the BHNS population, we expect that these signatures for BH spin can help to distinguish these two formation pathways --we note that other formation pathways are certainly possible for isolated binaries, see e.g., \citet{Broekgaarden2021}. 
To do so with statistical rigor would require hierarchical Bayesian analysis of BHNS binaries with gravitational-wave detections in conjunction with electromagnetic follow-up of kilonova observations, and more sophisticated models of binary evolution than considered here. Nevertheless, acquiring a statistically large population of observed BHNSs may prove difficult, especially ones with kilonova counterparts, implying that Bayesian analysis of individual events may also be useful. Although such a task is not trivial, it is certainly feasible since constraints on $f_{\rm a}$ have already been placed on the LIGO/Virgo BH binary events via a backwards sampling scheme and the COSMIC population synthesis code \citep{Wong2022}. 
For example, Bayesian analysis using gravitational wave data of a BHNS event and a detailed population synthesis model can provide constraints on $f_{\rm B}$, $f_{\rm a}$, and $m_{\rm BH}$, a detailed kilonova model used on electromagnetic observations can provide constraints on $m_{\rm ejecta}$, and together the possible formation history of the binary can be probed by leveraging the signatures identified here, e.g., Fig's~\ref{F:ContoursFaFb} and \ref{F:ContoursMassFracs}.

The strength of core-envelope coupling is an underlying process of the progenitors of stellar-mass BHs, implying this uncertainty affects all formation channels of BHNS binaries. When we showed the effect of accretion in Pathway B1, we assumed that core-envelope coupling was strong which provides negligible natal BH spin magnitudes. Binaries that evolve in Pathway A1 under strong core-envelope coupling will retain negligible BH spins in the absence of other spin-up mechanisms, which disallows significant $m_{\rm ejecta}$. 
In reality, the strength of core-envelope coupling is possibly not extremely weak or strong and depends on the mechanism that drives angular momentum transport within the stellar interior. Population synthesis models typically assume that core-envelope coupling is strong, however this is uncertain for high mass stellar progenitors of BHs \citep{Bowman2020}. We contend that better understanding of this process is crucial for predicting $m_{\rm ejecta}$ and the detectability of BHNS binary counterparts. If strong core-envelope coupling is prevalent in high-mass stars in nature, the results of this study offer predictions that can be used to rule out the presence of weak core-envelope coupling in the progenitors of BHNSs.

In our model, disk accretion can produce a highly spinning BH in Pathway B1 where the secondary star undergoes stable mass transfer. It is suspected that this accretion needs to be highly super-Eddington to achieve large BH spin \citep{Zevin2022}. Eddington-limited accretion would greatly suppress $\chi_1$, $m_{\rm ejecta}$, and the observability of any counterpart. However, super-Eddington accretion is not impossible in principle, as the Eddington limit depends on the geometry of the accretion and various kinds of instabilities. It is suspected that ultra-luminous X-ray binaries contain NSs accreting far above the Eddington limit, and some may contain accreting BHs that exceed the Eddington limit \citep{Miller2019,Gao2022}.

We also computed realistic light curves of kilonova counterparts to our BHNS binaries. Although there exist considerable uncertainties in the physics of kilonova, the light curve model utilized here is robust and reflects the current understanding. We showed that the kilonova emission that results from our highly spinning BHNSs are undetectable by ZTF, but will be discovered by future telescopes such as the Vera Rubin Observatory. Binaries that inherit high BH spin with $f_{\rm B} \gtrsim 0.03$ and binary mass ratio $Q = 3.6$ are predicted to produce kilonovae with peak brightness $M_i \sim -14.5$ for a few days of observing time and should be detectable from up to $\sim$500\,Mpc away in a 30s visit by Rubin.

Such systems may also produce short GRBs \citep[e.g.][]{Paschalidis15}, and we showed that the expected jet luminosity is consistent with the observed short GRB population. However, drawing direct links between the properties of the binaries and the GRB light curves is not possible due to uncertainties in the jet launch mechanism, the process by which $\gamma$-rays are produced in the jet, and the highly variable circumstellar environment. The isotropic nature of kilonovae also makes them more promising electromagnetic counterparts for gravitational wave detections of BHNSs compared to the strongly beamed short GRBs.

Although the two BHNS mergers GW200115 and GW200105 found by LIGO/Virgo had unfavourable parameters for producing electromagnetic counterparts, our results show that rapidly rotating BHs in BHNS binaries, significant ejecta masses, and detectable kilonova emission are possible through accretion and inheritance of spin. Comparing the distributions of ejected masses from future electromagnetic searches with the binary parameters inferred by gravitational wave detectors offers a promising means to determine the physical mechanism producing the BH spin in these systems.

\section*{Acknowledgements}
The authors would like to thank Christopher Berry for insightful discussion, and Daria Gangardt, Davide Gerosa and the Referee for very helpful comments. N.S. is supported by the Leverhulme Trust Grant No. RPG-2019-350, the European Union's H2020 ERC Starting Grant No. 945155--GWmining, and the Cariplo Foundation Grant No. 2021-0555. B.G. and M.N. are supported by the European Research Council (ERC) under the European Union’s Horizon 2020 research and innovation programme (grant agreement No.~948381). M.N. acknowledges a fellowship from the Alan Turing Institute.

\section*{Data Availability}
The data underlying this article will be shared on reasonable request to the correspondence author.

\bibliographystyle{mnras}
\bibliography{bibme}{}

\end{document}